\documentclass{svproc}
\usepackage{graphicx} 
\usepackage{natbib}
\usepackage{amsmath}
\usepackage{geometry}
\usepackage{url}
\usepackage{url}

\begin{document}
\mainmatter              
\title{Navigating the Evaluation Funnel to Optimize Iteration Speed for Recommender Systems}
\titlerunning{Navigating the Evaluation Funnel to Optimize Iteration Speed for Recommendation systems}  

\author{Claire Schultzberg \and Brammert Ottens\inst{1}}

%

%
\tocauthor{Claire Schultzberg, Brammert Ottens}
\institute{Spotify, Search, Stockholm, Sweden}

\maketitle              

\begin{abstract}
Over the last decades has emerged a rich literature on the evaluation of recommendation systems. However, less is written about how to efficiently combine different evaluation methods from this rich field into a single efficient evaluation funnel.
In this paper we aim to build intuition for how to choose evaluation methods, by presenting a novel framework that simplifies the reasoning around the evaluation funnel for a recommendation system. Our contribution is twofold. First we present our framework for how to decompose the definition of success to construct efficient evaluation funnels, focusing on how to identify and discard non-successful iterations quickly. We show that decomposing the definition of success into smaller necessary criteria for success enables early identification of non-successful ideas. Second, we give an overview of the most common and useful evaluation methods, discuss their pros and cons, and how they fit into, and complement each other in, the evaluation process. We go through so-called offline and online evaluation methods such as counterfactual logging, validation, verification, A/B testing, and interleaving.  
The paper concludes with some general discussion and advice on how to design an efficient evaluation process for recommender systems. 
\keywords{recommender systems, evaluation, offline evaluation, A/B testing, interleaving, information retrieval}
\end{abstract}

\section{Introduction}
Product evaluation is an important and natural part of modern product development. During the last decades, the literature on how to evaluate product changes in tech-products has exploded at the same rate as the industry itself. Machine learning-based products, including recommendation systems, are no exceptions. There is by now a large body of tools and methods for evaluating various aspects of machine learning-based products, see for example \cite{joachims2002optimizing, joachims2003evaluating, radlinski2008does, brochu2010tutorial, schuth2014multileaved, schuth2015predicting, chuklin2015comparative, stone2022book, diaz2022offline, bi2023interleaved,  faggioli2023perspectives, thomas2023large}. There are few overview papers available for this field. Some exceptions are, e.g., \cite{beel2015comparison} compared offline evaluation with online evaluation and user study in the context of research-paper recommender systems. \cite{garcin2014offline} compared offline and online evaluation for news recommendation. However, less has been written about how different online and offline methods compare and complement each other, and how several evaluation methods can be combined into one evaluation process.

In this paper we give an overview of important evaluation methods and propose an evaluation framework that aims at helping practitioners navigate the role of these methods and how they can be fitted together into an efficient machinery for ML product evaluation. 
Throughout the paper, we will use a personalized search system as the main example, but the proposed framework can be applied to recommender systems in general. 
While many may associate 'evaluation' with quantifying the impact of a product change, evaluation is also a powerful tool for learning and iterating in product development, allowing to improve the end-user experience and drive business value faster.

The rest of this paper is organized as follows. Section \ref{sec:eval_as_a_tool} explains how product evaluation and speed of iteration are related, and proposes ways to decompose the evaluation of a product variant that enables faster iterations. Section \ref{sec:the_eval_framwork} presents our evaluation framework. Section \ref{sec:offline_eval} gives an overview of offline evaluation methods, i.e. evaluation that can be done on historical data, before exposing real users to a new version. Section \ref{sec:online_eval} gives and overview of online evaluation methods, focusing on how they can be used to increase speed of iteration. Section \ref{sec:tradeoff} discusses trade-offs between doing certain parts of the evaluation offline and online, respectively. Section \ref{sec:summary} presents a summary and concluding remarks.

\section{Evaluation as a tool for iterating fast}\label{sec:eval_as_a_tool}

To iterate fast on any product, we must accept a few hard truths. The most important one is that most of our ideas will not be successful. Several big successful tech companies report that somewhere between 10-20\% of their ideas are successful \cite{thomke2020experimentation}. In the light of this sobering fact, the natural consequence is that to progress fast it is essential to minimize the time spent on non-successful iterations of our products. The main principle of the evaluation framework proposed in this paper is centered around precisely this: The sooner we can establish that an iteration is not successful \footnote{We will be more precise about what that means below.} and discard it, the faster we can move on to the next iteration and thereby find and ship the successful ideas. The good news is that finding necessary evidence for that an idea is non-successful is often much easier than finding sufficient evidence for that it is successful. In the next section we present ways to decompose success to enable early identification of non-successful iterations. Note that we say 'non-successful' rather than bad, when talking about ideas and iterations. This distinction is important, as it might be the implementation of the idea, rather than the idea itself that is making it non-successful. By using evaluation in a systematic way, we can also establish feedback loops for iterating fast on the implementation of a given idea.

\subsection{Useful ways to decompose the definition of success to speed up product iteration}
When the goal is to discard non-successful ideas as soon as possible, it is crucial to be precise about what 'successful' means. Of course, what the exact definition of success looks like will depend on the context and the goal of the current iteration of the product. Typically we start from a problem. For example for a search system it can be that some types of queries are less successful than average. We then build an hypothesis in the form of "To solve this problem/reach this goal we will implement this change/idea. We will be successful if we observe this outcome". 

We propose decomposing the definition of success into necessary and sufficient criteria. A necessary success criterion is something that the product change has to achieve to be regarded as successful, but achieving it is not enough to imply success. In other words, all successful product changes fulfill the necessary success criteria, but not all product changes that fulfill the necessary success criteria are successful. On the other hand, a sufficient criterion is something that, if achieved, directly implies success. A common sufficient criteria is a successful A/B-test (which often contains several necessary criteria in its analysis). The framework presented in this paper is centered around finding necessary criteria that can be evaluated as soon in the product development funnel as possible, even before conducting an A/B-test. For instance, in a personalized search system where users can search for music and podcast content, a version that returns only music content for all queries and users would never be considered successful. Detecting such issues early allows for quickly dropping that version and iterating to the next version. The more carefully we can observe the output of the system we are evaluating, the more precise and relevant we can make the necessary criteria. As we will see below, the concept of necessary criteria helps a lot when it comes to finding early-in-the-product-development sanity checks.

Another useful way of decomposing the evaluation is to separate \textit{verification} from \textit{validation} as first proposed in \cite{boehm1984verifying}. Boehm described these concepts as 
\begin{itemize}
    \item Verification: Are we building the product right?
    \item Validation: Are we building the right product?
\end{itemize}
\textit{Verification} is common practice in software development. For example, when testing different colors of a button in an A/B test, it is common to verify that indeed the color of the button is the intended one for all test groups before launching the test. However, verification becomes much more central to the evaluation funnel of recommendation products and in particular if they are personalized. This is because often the recommender systems are partly a black-box by construction, and if the system is complex, its interaction with other surrounding systems can make it difficult to predict how changes will propagate into the served experience. \textit{Validation} \footnote{Note that the validation that we talk about here is different from the validation step often done during ML-training.} on the other hand, is about evaluating the impact of the change, i.e., how users react to the change. We \textit{validate} that the change had the intended effect on the outcome defined in our hypothesis. 
In summary, when we make a product change, we can first \textit{verify} that the change is having the intended effect on the product, and then we \textit{validate} that the change has the intended impact on users according to our hypothesis.

The concepts of verification and necessary success criteria are related because verification can, and in our opinion should, be used as a necessary criterion. This is inline with Boehm, who said that about verification for software development '\textit{These recommendation provide a good starting point for identifying and resolving software problems early in life cycle-when they're s relatively easy to handle}' \citep{boehm1984verifying}. While various other necessary criteria exist, verification is perhaps the most natural one; if the implementation of the change is not changing the product as intended -- little suggests it will improve the product accordingly.

\section{A product evaluation framework}\label{sec:the_eval_framwork}
This framework presents a way of reasoning about evaluation and the role of different methods, in particular for recommendation systems. We see the product evaluation work as a continuous process and think about it as a funnel that each idea has to go through to be proven successful and thereafter shipped in production. The sooner in the funnel we can identify that an idea is not fulfilling a necessary criterion, the sooner we can move on to the next iteration. For people used to relying only on A/B tests, it might be useful to think about our framework as extending the evaluation process backward from the point at which an A/B test is run, to as early as possible in the product development. This way, we can avoid spending time and resources on running A/B tests for ideas that don't fulfill all necessary criteria that can be checked before running the test. This also means lowering the risk of exposing end users to bad experiences. 

For any product change to be regarded as successful, evidence for a number of things typically has to be in place. Generally, it is good to have established the following
\begin{enumerate}
    \item Verification that the change to the product is working as intended, i.e., is changing what it aimed to change.
    \item Quality assurance, i.e., that the change is not deteriorating the quality of the app/experience (load time, crash rates, etc).
    \item Validation that the change is causing the improved outcome it intended. E.g., more satisfied customers, higher click-through-rate, etc.
\end{enumerate}
Evidence for these things can generally be collected both during the offline and the online phase. However, there are trade-offs and limitations that make some things more natural to evaluate during the offline and online phases, respectively. For more details, see section \ref{sec:tradeoff}. We say that to establish success it is \textit{sufficient} that all of the above-mentioned criteria are fulfilled.

\begin{figure}[hbt!]
    \centering
    \includegraphics[width=\textwidth]{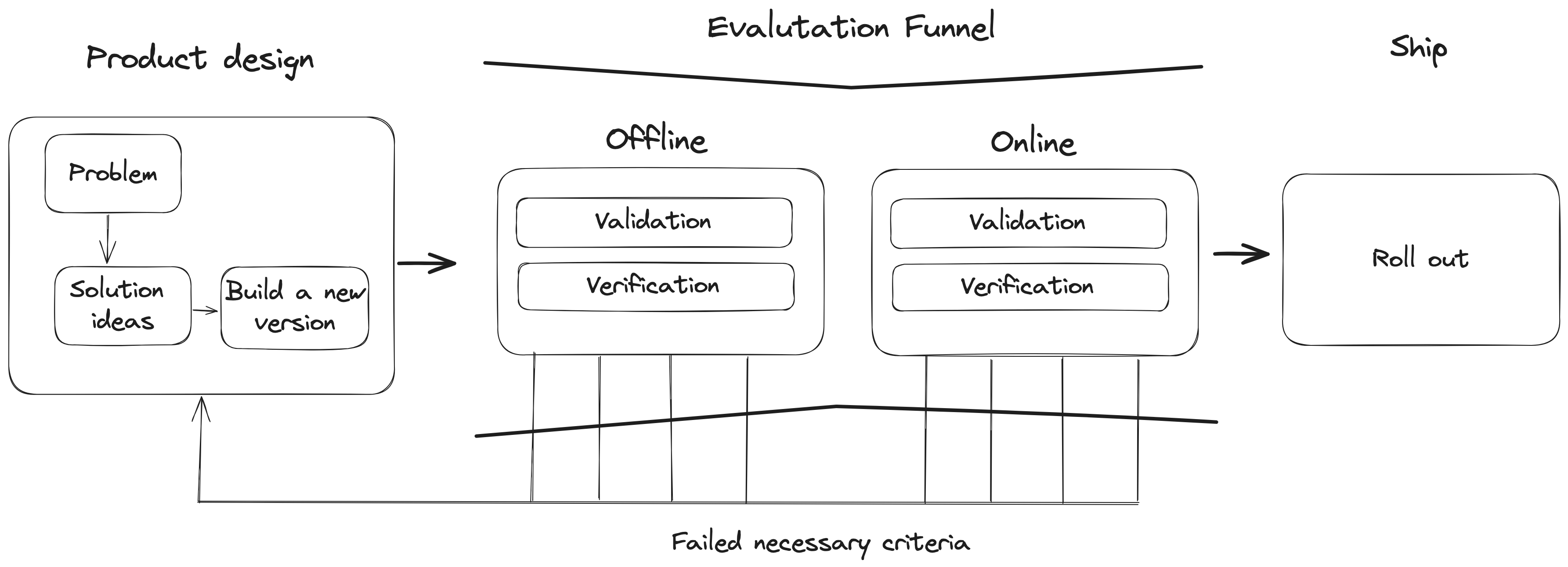}
    \caption{Schematic view of the evaluation framework.}
    \label{fig:schematic_flow}
\end{figure}

Figure \ref{fig:schematic_flow} displays a schematic view of the evaluation framework. There is a product design stage where the problem that we aim to solve in this iteration is defined and various ideas for solving the problem are discussed, which is consolidated into an hypothesis which is then implemented as a new version of the product (or several). This version is then fed into the evaluation funnel. The evaluation funnel has two parts, offline evaluation and online evaluation, and both can contain validation and verification. The key insight of our framework is that it is possible and desirable to find shortcuts out of the funnel that go immediately back to the design stage, where a failed necessary criterion is an example of such a shortcut. To build an efficient iteration-loop, it is important to both identify non-successful ideas early and understand why they failed.
The aim of this paper is to provide a light but rich overview of the options that are available for each step of the evaluation funnel, and to make their pros and cons clear. In the following sections, we look in detail at all the parts of the evaluation funnel as presented in Figure \ref{fig:schematic_flow}. 

\section{Offline evaluation}\label{sec:offline_eval}
For many recommendation systems, it is possible to perform part of the evaluation offline. Offline here means that we can evaluate a new version of a system without exposing any users to it, using historical data. It is possible to perform both verification and validation offline, albeit offline validation has a few major caveats that will be discussed below.

Before digging into the offline evaluation methods, we need to build some intuition for the concept of counterfactual reconstruction which is the mechanism that enables most offline evaluation. 

\subsection{Counterfactual reconstruction and practical considerations}
The key word in counterfactual reconstruction is \textit{counterfactual}, i.e. it is about the "what would have happened". In the context of this paper, by counterfactual reconstruction we mean logging all the information needed to reconstruct offline the recommendation a user would have seen for a given query if model X was used. An alternative to counterfactual reconstruction is so-called \textit{counterfactual logging}, which means logging the results returned by several models when a given query comes in. Some model X might be the model that serves the results in production, but we can (counterfactually) log what models Y and Z would have returned at the same time for the same query and user. 

When implemented properly, counterfactual reconstruction is more flexible than counterfactual logging as it allows to evaluate any new model before is even has to be productionalized. For example for a search system, for a given user and query this would mean logging all the features for the user, query and for all the retrieved candidates, so that any model can be used offline to reconstruct the ranking of the results for that user and that query. It can be a powerful tool that enables us to learn a lot about how the end-user experience is affected by the product change we want to evaluate, before anyone has been exposed to it. On the other hand, in complex systems where many parts are working together to provide the recommendation (like modern search stacks), it is not trivial to obtain everything that is required for counterfactual reconstruction without errors. This is because in many systems, a model doesn't operate in isolation. There will be other models, heuristics and business rules that all interact to generate the results the user will be presented with. An additional difficulty is that many things change over time, e.g. models are retrained, content is coming and going, etc, a logging system for reconstruction as described above can be hard to construct without losing quality as compared to counterfactual logging.

A crucial aspect of counterfactual reconstruction is obviously that it allows to accurately reconstruct what the users have seen and would have seen under the alternative. Fortunately, it is possible to quality-test counterfactual reconstruction (and logging) by reconstructing (or logging) the results for real queries using the model that is in production and comparing them with the results actually returned for the same query.
For counterfactual reconstruction to be of high quality the following behaviour is required
\begin{itemize}
    \item If the results from the same model are logged in production and with counterfactual reconstruction (or logging), they should be the same.
    \item If we repeat the counterfactual reconstruction for a given model and a given input, the results should be the same each time.
\end{itemize}
Two useful measures to quantify similarity in recommendations are the Jaccard Index \citep{jaccard1912distribution}, and Spearman's Rank Correlation Coefficient \citep{spearman1961proof}. The Jaccard index only quantifies the overlap in terms of content, whereas Spearman's Rank Correlation Coefficient also takes the position/rank of the content into account.
How similar to production the counterfactual logs need to be as measured by e.g. Jaccard index or Spearman rank correlation coefficient depends on the use case, but clearly the value of using counterfactual reconstruction diminishes quickly if the discrepancy becomes substantial.  

Now that counterfactual reconstruction is well defined, we are ready to dig into the offline verification methods. 

\subsection{Offline verification}
Offline verification offers a lot of possibilities. Here we will focus on the possibility of finding necessary criteria, but offline verification also offers a rich feedback loop about how a change to the system impacts the recommendations served to users. 
We will look closer at a few of the most interesting themes of insights that can be learned in the offline verification step. 
\begin{itemize}
    \item \textbf{Width versus depth} -- for how many queries the results changed (width), and how much did the query results that changed change (depth)? This gives immediate information about how the change is realized in the search system. For depth, the Jaccard Index and the Spearman's rank correlation are again useful measures. 
    \item \textbf{Width vs depth segmented by success} -- For the results given to a user in production, we often have a notion of success, e.g., proxied by clicks or consumption of an item. We can look at the width and depth of the change for the previously successful and unsuccessful queries.
    \item \textbf{Width vs depth segmented on query characteristics related to the product change} -- if we intended to improve the results for long queries, we can ensure that it was mainly long queries that got different results.
\end{itemize}

Many necessary criteria can be constructed based on the information from these three simple descriptive statistics. Some examples of generally applicable necessary criteria are
\begin{itemize}
    \item  Did sufficiently many results change sufficiently much to hope for impact on user experience?
    \item  Did responses to sufficiently many previously non-successful queries change?
    \item  Did the results to too many previously successful queries change?
    \item  Did results change enough (width and depth) for the types of queries the product change aimed for? 
\end{itemize}
There are of course situations where some of these checks do not make sense. For example, if the search is mostly successful already, the aim might be to improve quality, in which case the second bullet might not apply. The takeaway is that for any given product change, it is possible to construct a set of verifications that can act as necessary criteria for success, which enables discarding non-successful ideas early in the funnel. 

The proportion of previously successful/non-successful queries that are affected also gives some information on the impact we can expect from the change. For example, consider that only the results of successful queries are affected. Then, the success rate can only be unchanged or decreased in such a scenario, which might make the value of pursuing the idea further questionable. On the other hand, if only previously non-successful queries are affected, there is little to lose in terms of success rate, but these queries are likely the most difficult to improve.

Note that, since offline verification is purely descriptive, no assumptions (other than high-quality counterfactual reconstruction) are necessary to trust these insights. This is not the case for offline validation, which often relies on strong assumptions, something we will return to in detail. 

\subsubsection{Human evaluation}
In the context of recommendation system evaluation, 'human evaluation' covers various approaches and methods and we will return to it in section \ref{sec:offline_val}. 
Human evaluation can be used to evaluate necessary criteria during the offline verification phase. For example, one necessary criterion can be that for a set of queries for which there is an obvious "right" result, that result should always be present and ranked high. These queries can then be run through a new model to verify if the results are indeed present and ranked sufficiently high. Human judges can create this set of queries and results. Note that in highly personalized systems it might not be trivial to come up with these queries and results. It is important to ensure that the current production model (also over time) is passing such criteria before adding them to the evaluation funnel.

Developing a human evaluation process can be complicated and costly, it requires to have access to the right resources and skills both in terms on human judges and internal resources to create the guidelines and train the judges \citep[~section 7.2]{stone2022book}. 

\subsection{Offline validation} \label{sec:offline_val}
As mentioned above, offline validation is not so much about determining whether a model behaves as we expect, but to validate that our new approach is more relevant to our users. In the context of search, for example, that means that our new approach is better at showing relevant candidates to the user. To answer this question offline we need to solve two sub problems. The first is how to determine the relevance of a candidate for a query, and the second is how to determine which of two result sets is better. For the first problem, there are different metrics available and we won’t spend much time on those in this paper. The main idea is that those metrics consider the relevant candidate previously determined as successful. See \cite{ canamares2020offline, gilotte2018offline, stone2022book} for good overviews.

For the second problem, there are historically speaking two general approaches. The first relies on historical click logs (or other success signal), i.e. the behaviour of actual users, while the second relies on dedicated human evaluators that judge either only the relevance, or evaluate the whole result page. More recently, we have also seen the use of Large Language Models (LLM) to provide relevance judgements, see~\cite{thomas2023large}. In our view, all three have their place in an offline evaluation pipeline.

\subsubsection{Click log based evaluation}
Using click logs for relevance judgements is a simple idea, and used throughout the industry. The interactions of (a random sample of) users/queries are logged and used to determine whether a result is relevant to the query. In search, for example, we can use a click on a result as a signal of success. We can also use richer information such as subsequent actions to determine success. There are a few drawbacks and assumptions with this method, however

\begin{enumerate}
    \item It assumes that the success signal used accurately represents the relevance of an item for a query.
    \item The user's behavior is biased by the position and the way the results are shown.
    \item There needs to be overlap between the results returned by the production and the new approach, otherwise there are no relevant judgements available for the new candidates.
\end{enumerate}

The problem of bias can, in many cases, be overcome by estimating this bias, and then controlling for it in the metrics used to score approaches \cite{diaz2022offline}.

The overlap issue however, is more difficult to overcome \citep{buckley2004retrieval}. Overlap is a term we have borrowed from the causal inference matching literature. 
Because we are using historical interactions, for offline validation to give useful insights there must be large \textit{overlap} both between the currently successful type of queries and the type of queries we intend to change with the new version and between the results returned by the two approaches.
For example, if we are making a change aiming to improve currently non-successful queries, then the offline verification phase should show that mostly non-successful queries changed and as a consequence, we do not know which item would be successful. Similarly, if we make a drastic change (in depth), the new system will return new unseen candidates for many results, candidates for which we don't have relevance judgment.

Human evaluation and LLMs, as discussed in the next sections, are good candidates to reduce the bias and overlap problems.

\subsubsection{Human evaluation}
One common practice is to present humans judges with queries and associated results and ask them to rate the relevance of each item according to guidelines \citep{stone2022book}, give preference for one set of result over another, or rate a full result slate for a query. In that context, human evaluation doesn't suffer from the bias or overlap problems of the click logs approach. 

Human evaluation relies on the assumption that trained judges can assess the relevance of a given document for a given query like the end user would \cite{stone2022book}. This assumption might not hold for personalized systems where the same query can be typed by several users with very different intent and different notions of what a relevant document will be.
Another drawback, as mentioned before, is that scaling human evaluation can be complicated and expensive, as opposed to scaling click log evaluation.

\subsubsection{LLM based evaluations}
With the introduction of LLMs, the opportunities to obtain relevance judgements at scale are increasing drastically. The idea is to replace a human judge with an LLM, either by providing a textual representation of the recommendation results, or even a graphical representation of the results, allowing the LLM to not only evaluate the candidates and the order in which they are presented, but also the way they are presented to the user, i.e. the font used, any artwork that is shown with the results, or which parts of the text are highlighted.

LLM's come with their own problems and drawbacks however. Little is known about the biases these models would introduce in the evaluation, and they need to be investigated more and thus used with care \citep{thomas2023large, faggioli2023perspectives}.

\section{Online evaluation}\label{sec:online_eval}
In online evaluation, both the current production version and the new version are exposed to real users. This makes it possible to estimate the causal effect of the change on any metrics we care about. 

\subsection{Online verification} 
As discussed above, verification can be efficiently performed offline, without exposing real users and without making assumptions on how users would react to the change.
A successful A/B test can often be considered a sufficient criterion. However, if one has to, or chooses to rely only on online evaluation, there is a case to be made for enabling verification as well. This is because, even if A/B-tests can provide sufficient evidence of success without verification, to maximize the learning of the non successful iterations, we want to understand if it was because the current iteration failed to realize the intended change to the system (verification), or if the intended change was not having the intended impact on the outcome. That is, for ideas that fail validation, verification can give a lot of insight for how to proceed with the next iteration. 

To perform verification online, counterfactual reconstruction is required. Since verification is a within-person-and-query evaluation, we need to generate the results under both systems for each query, regardless of which system actually provided the results to the user. 

In our view, there is little to be gained from doing verification online rather than offline. There are reasons to implement counterfactual reconstruction regardless (we will get back to this in section \ref{sec:exposure}), and if counterfactual reconstruction is in place we might as well save time and discard non-successful ideas safely, already in the offline stage.

\subsection{Online validation}
We argue that the only way to validate recommendation and search systems without making strong assumptions is via online validation.
This is because the reward often relies on implicit signals such as clicks, purchase or consumption time for an item that the user chooses among others \citep{castells2022offline}. For this reason, it is crucial to expose users to the (new) alternative set of documents for them to choose from.

The most common way of performing online validation of a product change is by using randomized experiments, also called 'A/B tests', where users are randomly assigned to the different variants of the system and we look at e.g. differences in averages (so called average treatment effects) on any metric of interest. A/B tests have been established as the gold standard evaluation for most product evaluation in tech for the last decades. We agree that A/B testing has a place in most evaluation funnels.
Since much is written about standard A/B-testing on recommendation systems and search, we will here instead focus on how to use A/B tests for fast iteration and how other more advanced methods such as interleaving, multi-armed bandits and Bayesian optimization can be leveraged as well. 

\subsubsection{Sequential testing in A/B tests}\label{sec:seq_test}
A lot of methodological development has been directed at speeding up the experimentation rate. That is, the speed with which we can learn from experiments. A central part of this work has been around so-called sequential testing. For recent state of the art sequential testing, see e.g. \cite{jennison1999group, howard2021time, waudby2021time}. Sequential testing is a family of confidence sequences and hypothesis tests that allows for evaluating the experiment during the data collection -- before the experiment is concluded. This allows for stopping the experiment as soon as a significant result is obtained. It can be shown that, in expectation over repeated experiments, sequential tests decrease the sample size as compared to standard methods. However, sequential tests also have a few limitations. One limitation of sequential tests is that they, due to the flexible stopping rules, yield biased treatment effect estimators, see e.g. \cite{fan2004conditional} for group sequential tests. That is, if experiments are stopped at the first significant result, the treatment effect will be overestimated in expectation, i.e. biased upwards. 

A second limitation is that given a certain sample size, sequential tests are more conservative than standard tests. This is natural since sequential tests are developed to correct the otherwise inflated false positive rate. However, this means that if experiments need to run for a certain time, or at least until a certain number of users have come into the experiment, the sequential tests become overly conservative.
There are strong reasons for why it is desirable in many product evaluation processes to run the experiments for both a fixed length and until at least a certain number of users have entered. The length of the experiment is important for products that have seasonality effects, i.e., that the treatment effect depends on e.g. the day of the week or that only very active users are entering the experiment early in the experiment \citep{wang2019heavy}. To obtain an unbiased estimator for the treatment effect averaged over the weekdays, the experiment needs to run for at least a week, such that users are observed across all the days. In addition, it is well known that under-powered experiments, i.e., experiments that have too low probability of detecting a true treatment effect, are problematic. Significant effects in under-powered experiments are highly likely to overestimate the true treatment effect \citep{gelman2014beyond}. This too points to that sequential tests with early stopping is problematic if the point estimate is of interest.

In summary, sequential tests are efficient at finding if there is a significant change due to the treatment (i.e., the new version) or not, but less good at quantifying the magnitude of that effect. For these reasons, many are now recommending using sequential tests during the experiment -- not to make shipping decisions -- but rather to monitor if the experiment is harming the systems or the end users, and if this is the case abort the experiment as soon as possible \citep{kohavi2012trustworthy}. The argument for why this is a good idea, is that biased estimators is less of a problem for detecting non-success, than for validating success. By using sequential tests for guardrail metrics and quality metrics like crash rates or load times, etc -- we obtain a shortcut to discarding non-successful ideas even during the A/B-test. 
This way of separating the decision to discard an idea versus the decision to ship a new version of a product is highly relevant for the framework introduced in this paper. We can define a necessary criterion by stating that for a product change to be successful, it should not compromise the user experience or the product quality and test this criteria early in the online evaluation phase in a statistically rigorous way by using sequential tests. See \cite{schultzberg2024riskaware} for more details on how to design experiments with several decision criteria.

\subsubsection{Variance reduction in A/B tests}
Other efforts have been made to improve the speed of learning and efficiency of experiments. Variance reduction is a set of statistical techniques using additional data on users in the experiment analysis to reduce the variability of the treatment effect estimators. The most popular method for variance reduction, is so called regression adjustment \citep{cochran1957analysis} popularized in online experiments by \cite{deng2013improving}. Regression adjustment has the nice property of ensuring the average treatment estimator is still unbiased, see e.g. \cite{negi2021revisiting} for a recent summary. In short, lower variability implies higher power, which in turn means that experiments with smaller sample size can yield the same power with variance reduction, as larger experiments can without variance reduction. Although variance reduction is a powerful and useful technique, it should in our opinion not be mixed up with 'faster' experiments in the sense that the length of the experiment needs to be planned taking more things than efficiency into account, as discussed in Section \ref{sec:seq_test}.

\subsubsection{Exposure filtering using counterfactual logging to gain precision in A/B tests} \label{sec:exposure}

A less popularized strategy for improving precision in experiment results is what is sometimes called 'exposure filtering'. The idea of exposure filtering is that, even though everyone in the experiment has been served the experiment version of the product, not all those users have actually seen or interacted with the actual change that is evaluated. For example, if we make a change to a certain page in an app, only the users who navigated to that page during the experiment have truly been exposed to the change. The idea with exposure filtering, is that only the users that actually were exposed to the change are included in the results \footnote{The users in the control group that navigated to this page might not see a change but are logged as exposed.}. Naturally, including users that were not exposed dilutes the treatment effect and can increase the variability of the estimator. This implies that filtering these users out improves precision. 

For personalized search and recommendation systems the notion of exposure becomes tricky since the results the users are exposed to depends on the query or user context. Therefore, we cannot easily determine which users are exposed or not because while we may know what recommendations users have seen, we don’t know if they would have seen different results under the alternative.
Counterfactual logging (or triggered logging) can be leveraged to increase the precision in online A/B tests  \citep{kohavi2012trustworthy,stone2022book}. The idea is simple: by filtering out the queries for which the two models return exactly the same results, we reduce the treatment effect dilution and variance of the treatment effect estimator leading to more efficient tests. 
It is possible to filter out users that had no queries for which the results were different between the models during the experiment, instead of doing it on the query level. However, since one user can make more than one query, many users will have at least one query that changed which disqualifies them from the filter, making the filter less impactful. On the other hand, filtering on the query level implies that query level metrics and treatment effect estimands have to be used which makes interpretation more complicated and relies on the assumption that observations are independent, which might not hold in practice.

In summary, the choice of filtering out based on counterfactual logging on the user level or the query level is a trade-off between efficiency, flexibility and interpretability in the validation step. 

Exposure filtering is somewhat related to the concept of so-called average treatment effect of the treated. We can view the exposed users as the 'takers of the treatment' in the traditional framework of compliers \citep{imbens2015causal}. This also means that the change in interpretation of the treatment effect estimator under exposure filtering, as compared to the average treatment effect, needs to be taken into account. Using exposure filtering will increase the chance of detecting if the new product version had an effect on anyone, but the average treatment effect estimator of the treated (exposed) does not translate into what the value of shipping the new version is. For example, say that 5\% of the users are truly exposed to the change and we detect a lift of 1\% among those exposed users. Then, the overall lift if this version is shipped will be somewhere around 5\%$\times$1\% = 0.05\%. In other words, it is important not to mix up the ability to detect effects, with relevant discoveries.       

\subsubsection{Interleaving}

\begin{figure}[hbt]
    \centering
    \includegraphics[width=\textwidth]{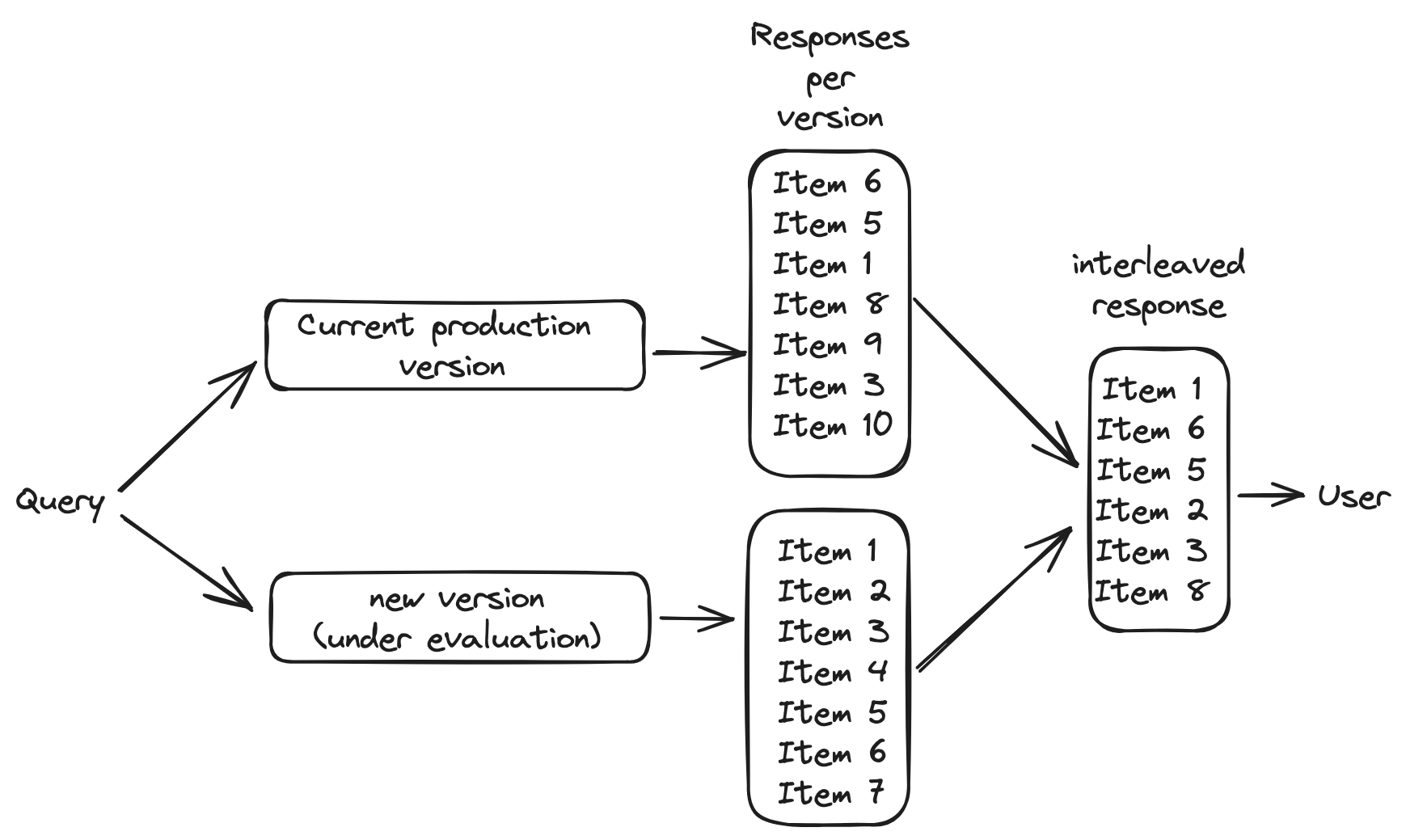}
    \caption{Simple flow for how a single interleaved result is constructed from results from two different versions of the recommendation.}
    \label{fig:interleavning_scheme}
\end{figure}

Besides trying to make traditional experiments (often called A/B tests) more efficient using sequential tests, variance reduction and exposure filtering, other evaluation methods have been proposed. For recommendation systems with retrieval and ranking. A method that has gained a lot of attention is interleaving \citep{joachims2002optimizing, joachims2003evaluating, radlinski2008does, schuth2014multileaved}. Interleaving (and multileaving) is a type of experiment, where instead of randomly assigning users to one version or the other, each users is exposed to several versions of the product simultaneously. A/B tests are by construction between-unit and therefore between-query evaluation. Interleaving is an alternative to A/B-testing that uses within-unit or within-query evaluation. In short, interleaving is a method where responses from both the current production model and the new model are semi-randomly assembled into a new \textit{interleaved} response. Figure \ref{fig:interleavning_scheme} displays the flow of interleaving. It is in the step when the responses from the two versions are combined into one that different types of interleaving algorithms can be used, see, e.g. \cite{chuklin2015comparative} for a comparison.

In its simplest form, once the interleaved result is constructed and served to the users, we can assess the within-users preference for the different models by giving scores to the model that produced the successful document. 

The within user comparison makes interleaving very efficient in comparison with A/B-tests \citep{bi2023interleaved}. For this reason, interleaving is sometimes presented simply as 'a more powerful alternative to A/B tests'. We would like to nuance this by pointing out that this is only true as long as preference, or at least a within-user metric, is the only validation objective. If any other type of metric is to be included as necessary criterion in the online validation, for example average lift in content consumption or financial metrics, interleaving is no longer an option that can replace A/B tests. An A/B test is much more flexible in terms of what treatment effect can be estimated. Although it is possible to use more sophisticated metrics than pure preference \citep{schuth2015predicting}, the type of metrics and estimands that can be used with interleaving is bound by the within-user comparison. 

A case can be made for using interleaving (or multileaving) as an online step before A/B-testing under certain conditions. If we typically come up with enough versions of the system to test that passes the offline necessary criteria for A/B tests to become the bottleneck -- and if superior preference is a necessary criterion for success -- we could use interleaving as a step before A/B testing to only A/B test the version that maximizes preference against control. 

\subsubsection{Multi-armed bandits}
A multi-armed bandit (MAB) is a (between-user) reinforcement learning method for finding and serving the best out of several variants, see for example \cite{slivkins2019introduction} for a comprehensive introduction. At a high level, a MAB is an algorithm that chooses how to distribute the users that come into the online validation to different treatment variants (arms) to minimize some regret, or maximize some utility. In the case of a recommendation system, for example, the arms could be versions of the ranking, and the utility function could be the proportion of successful queries. The algorithm goes through several rounds, where it uses the current information about the success of each arm (variant of the recommendation system), to inform how to distribute the user across the arms for the next round. MABs are appealing when there any many variants and we want to use online evaluation to quickly sort out which alternative is the best while at the same time limit the exposure of the worst versions. 

MABs are similar to interleaving in the sense that they reduce the risk of having a variant that is potentially bad for the end users in the online evaluation phase. An arm that performs poorly in the MAB will quickly be served to very few users (or dropped depending on the algorithm) thereby limiting the negative impact. As opposed to interleaving, MABs offer between-user comparisons, making it possible to use more standard business metrics for optimization and evaluation. In fact, MABs are similar to multivariate A/B tests and it is possible to use them in a similar fashion to make inference. However, due to the treatment (arm) assignment being a function of the outcome data, unbiased estimation of treatment effects\citep{dimakopoulou2021online, woong2023design} and sequential testing \citep{liang2023experimental} require special attention for MABs.

\subsubsection{Bayesian Optimization}
Another interesting method is Bayesian Optimization (BO), see e.g. \cite{frazier2018tutorial} for a introduction. BO is a crucial technique in machine learning, especially for optimizing complex and expensive objective functions \citep{brochu2010tutorial}. BO is perhaps most commonly used in (offline) hyperparameter tuning, where the goal is to find optimal settings for a machine learning model. 

In an online validation context, BO can be viewed as a modern version of the traditional surface response experimental design \citep{box1992experimental}. The classic example of surface response design is to try to find the right dosage of a medicine to maximize some utility (health), measured by the treatment effect. However, one can think about surface response design as a general optimization problem. In search, it is not uncommon to have a set of parameters e.g. number of results shown, share of different type of documents in the result, etc. where we are interested in finding the combination of parameter values that maximizes the improvement (treatment effect). A naive approach is to construct a hyper grid of combinations of parameter values and run multivariate A/B tests or MABs to find the best value. But in cases where the parameter space is of several dimensions and/or the parameters can take many values, this quickly becomes too inefficient and simply implausible \citep{shahriari2015taking}. Bayesian optimization uses Gaussian Processes to learn the response surface iteratively. We start with trying a few semi-random parameter value combinations, and then fit a Gaussian Process to predict what parameters will maximize the utility, and what parameters to test next to maximize the learning about the relation between the parameter space and the utility. This way, BO efficiently finds good combinations in relatively few iterations.   

Naturally, BO can be applied both offline and online. In an offline setting, BO can be used in a loop using offline verification or validation methods to calculate the utility function. Say, e.g, a search team is trying to find a good set of parameters to boost a certain kind of document in the search results to a certain proportion. Using counterfactual logging and verification, it is straightforward to try a few settings, and use Bayesian optimization on the proportion of the document type observed in the offline verification, where for example the squared difference between the desired proportion and the observed proportion can be used as a minimization function. The BO framework can then propose new parameter combinations, and this can go on iteratively until a set of parameters are found that gives a proportion of the document close enough to the target. 

The same thing can be done in an online evaluation step with the treatment effect as an outcome to maximize. In this case each new parameter setting (or batch of settings) are evaluated in a new A/B test. In other words, if iterations of the product are parametric, BO offers an efficient way to traverse the parameter space for good or even optimal settings.

\section{Trade-off between offline and online evaluation}\label{sec:tradeoff}
As hopefully is clear from this paper, there is a lot to gain from splitting the evaluation of our necessary criteria between offline and online. It is of course tempting to move as much of the evaluation as early in the funnel as possible, i.e., do most of the evaluation offline and never have to bother with A/B-tests and other online methods. At the same time, since a successful A/B test is often a sufficient criterion for success, it might feel tempting to skip all the extra work of offline evaluation and simply run A/B tests as fast as possible. 

In our opinion, verification offers enough insights to be valuable in itself. That is, 
understanding what changed (and not) in the responses due to the change that the new version implies to the system is key insight for future iterations. This is true both for successful and non-successful iterations. 

When it comes to validation, we think that it is important to have reasonable expectations on the offline validations \citep{castells2022offline, garcin2014offline, beel2015comparison}. As described in Section \ref{sec:offline_val}, many conditions have to be met for offline validation to give solid \textit{sufficient} evidence for success. In fact, the limitations of offline validation, with or without human evaluation, are so substantial that we think it best to see offline validation primarily as a set of necessary criteria that can help detecting non-successful ideas in the offline stage. When it comes to complete validation whether the change has the intended effect, we argue that there will be few situations where the offline validation is sufficiently rich and trustworthy. In addition, we are often interested in validating on metrics beyond those supported by offline methods. Add to that, that if offline evaluation (in particular verification) is done right, fewer ideas (relative to without it) should make it to an A/B test, and for those ideas it might be worthwhile getting the full set of insights that can be learned from an online A/B test. 

\section{Summary}\label{sec:summary}
In this paper we introduce a product evaluation framework that enables fast product iteration by identifying and discarding non-successful ideas early in the development cycle. We show how the definition of success can be decomposed into necessary and sufficient criteria, and formulate the evaluation process as a funnel with an offline and an online phase where \textit{verification} and \textit{validation} can be performed to evaluate the criteria. 
We give an overview of the rich field of evaluation methods focusing on how each method can contribute to speed of iteration in various stages of the evaluation funnel, and their pros and cons.

We discussed that implementing counterfactual reconstruction is almost always recommended for evaluation of recommender systems. Counterfactual reconstruction enables knowing what results a certain request would have returned, if it been given to another version (or variant) of the product than the version it was in fact given to. This creates the possibility to understand how different versions of the systems differ in their results to the same requests. Counterfactual reconstruction and logging are central to both offline and online evaluation, and is helpful both to speed up iterations, and help the product teams understand the evaluation results. 

A natural first step in the evaluation process is verification. Given that counterfactual reconstruction is in place, verification can be done offline, without making strong assumptions, and thereby allows for detecting non-successful implementations before exposing any users.
The verification step helps \textit{verify} whether the implementation of the current iteration cause the intended change to the product. It is particularly crucial for machine learning-based products such as personalized recommendation or search where the results displayed to each users can be unique to that user.  

Once we have verified that the current product iteration changes the output in the intended way we can be confident in our implementation and move on to validation. 
With validation we \textit{validate} whether the users behaviour changes according to our hypothesis, e.g., users become more satisfied, consume more, or retain longer. Recommendation and search systems attempt to present candidates from which the user selects the (most) relevant item by clicking on, or consuming, it. Performing validation offline, i.e., without exposing users to the new variant, relies on assumptions that can be hard to meet in practice. However, many of these assumptions can be at least fulfilled using a combination of click logs, human evaluation and LLM based evaluation to get relevance judgements. 
Doing validation online can be done without assumptions by exposing users to the different versions of the system and observing user behaviour directly. The most common way of evaluating online is by running A/B tests. In that case it is possible to detect non-successful ideas early (in the online evaluation phase) by using sequential testing on guardrail and quality metrics and abort the test early if harm is detected. It is also possible to make the A/B test more sensitive by using variance reduction and exposure filters, for example based on counterfactual logging.
In the case where the speed of iteration is limited by the large number of versions that pass the verification, it is possible to run multivariate A/B tests, but more efficient methods such as interleaving, multi-armed bandit and Bayesian optimization can be leveraged as well. 

When designing a product evaluation process for a given product, which methods will be most useful depend on the definition of success for that product and which necessary and sufficient criteria that definition can be decomposed into. 
If the definition of success is complex and can be decomposed into many necessary criteria, it is often beneficial to utilize several methods both offline and online to identify and discard non-successful iterations fast. If on the other hand the definition of success is simple, using, e.g., only online validation via A/B testing might be the fastest approach. How motivated it is to have a highly sophisticated evaluation funnel with both offline and online evaluation steps might also depend on the cost of the development cycle in terms of time and resources. The framework presented in this paper aims to help ML-product teams to reason about what evaluation process and tools make most sense given the requirements and circumstances of their product.

%
%

\bibliography{library}

\end{document}